\documentclass[sigconf]{acmart}
\usepackage{graphicx} 
\usepackage[caption=false]{subfig} 
\usepackage{adjustbox}
\usepackage{colortbl}  
\usepackage{amsmath} 
\usepackage{enumitem} 
\AtBeginDocument{%
  }

\copyrightyear{2025}
\acmYear{2025}
\setcopyright{cc}
\setcctype{by}
\acmConference[ICMR '25]{Proceedings of the 2025 International Conference on Multimedia Retrieval}{June 30-July 3, 2025}{Chicago, IL, USA}
\acmBooktitle{Proceedings of the 2025 International Conference on Multimedia Retrieval (ICMR '25), June 30-July 3, 2025, Chicago, IL, USA}
\acmDOI{10.1145/3731715.3734426}
\acmISBN{979-8-4007-1877-9/2025/06}
\begin{document}

\title{Audio-Visual Driven Compression for Low-Bitrate Talking Head Videos}



\author{Riku Takahashi}
\affiliation{%
  \institution{Hosei University}
  \city{Tokyo}
  \country{Japan}}
\email{riku.takahashi.4q@stu.hosei.ac.jp}

\author{Ryugo Morita}
\affiliation{%
  \institution{Hosei University}
  \city{Tokyo}
  \country{Japan}}
\email{ryugo.morita.7f@stu.hosei.ac.jp}

\author{Jinjia Zhou}
\affiliation{%
  \institution{Hosei University}
  \city{Tokyo}
  \country{Japan}}
\email{zhou@hosei.ac.jp}


\begin{abstract}
Talking head video compression has advanced with neural rendering and keypoint-based methods, but challenges remain, especially at low bit rates, including handling large head movements, suboptimal lip synchronization, and distorted facial reconstructions.
To address these problems, we propose a novel audio-visual driven video codec that integrates compact 3D motion features and audio signals.
This approach robustly models significant head rotations and aligns lip movements with speech, improving both compression efficiency and reconstruction quality.
Experiments on the CelebV-HQ dataset show that our method reduces bitrate by 22\% compared to VVC and by 8.5\% over state-of-the-art learning-based codec. Furthermore, it provides superior lip-sync accuracy and visual fidelity at comparable bitrates, highlighting its effectiveness in bandwidth-constrained scenarios.
\end{abstract}

\ccsdesc[500]{Information systems~Multimedia streaming}

\keywords{Audio-Visual Video Compression; Lip Synchronization; Low-Bitrate Codec}


\maketitle

\begin{figure*}[t]
    \centering
    \includegraphics[width=\textwidth]{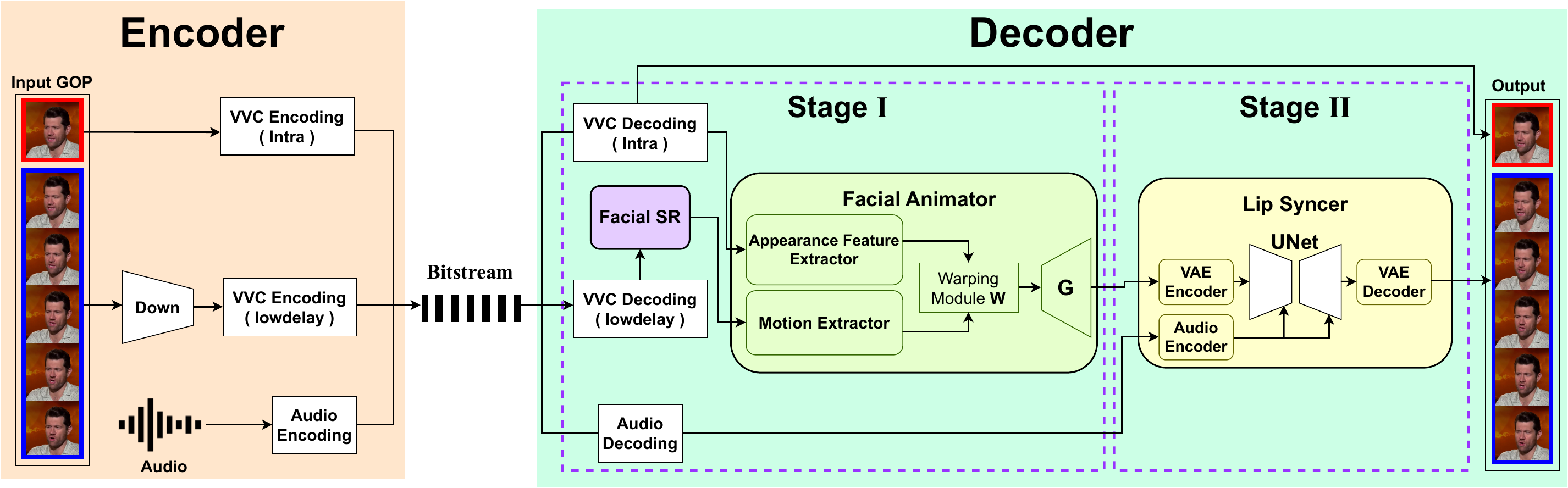}
    \caption{Architecture of the proposed codec. The images enclosed in red boxes represent keyframes, while the images enclosed in blue boxes represent target frames. The video reconstruction process at the decoder is divided into two stages. The first stage utilizes visual features to temporally reconstruct head movements and facial expressions. The second stage utilizes audio features to reconstruct the final frames where the audio and lip movements are synchronized.}
    \label{fig:architecture}
\end{figure*}

\section{Introduction}
The global pandemic has accelerated the adoption of video conferencing, making real-time audio-visual communication indispensable across diverse fields such as remote education, telemedicine, business meetings, and social gatherings. In these scenarios, the efficiency and quality of video compression are critical to ensuring seamless user experiences, especially under bandwidth constraints. 
HEVC~\cite{sullivan2012overview} and VVC~\cite{bross2021overview}, renowned for their high compression efficiency, have been widely adopted as conventional and advanced approaches for video compression.
In recent years, learning-based codecs ~\cite{konuko2021ultra, konuko2022hybrid, chen2022beyond, konuko2023predictive, 10992457} have demonstrated superior performance compared to conventional approaches. In particular, they have achieved high-quality reconstruction under challenging low-bitrate conditions, which was difficult for traditional methods, thereby improving encoding efficiency.

Despite remarkable progress, existing conventional and learning-based codecs have overlooked the crucial role of audio in guiding facial reconstruction. In reality, audio signals strongly correlate with facial shape and lip movements, providing valuable, integrated cues that improve talking-head synthesis. However, this correlation is often ignored, leading to misaligned lip shapes and suboptimal visual quality. Even slight desynchronization between a speaker’s lips and voice introduces perceptual discomfort, ultimately undermining communication effectiveness. 
Moreover, most learning-based video codecs predominantly rely on 2D keypoints to reconstruct faces across frames.
While these keypoints effectively represent local facial dynamics, they often fail to capture significant head rotations or substantial variations in viewpoint, resulting in distorted or inconsistent facial reconstructions.

To address these problems, we introduce a novel learning-based audio-visual driven video codec that combines the strengths of compact facial representations and audio-driven alignment to maintain high compression ratios without sacrificing perceptual quality. 
First, we introduce compact 3D features, enabling robust modeling of large head movements and complex facial orientations. 
These 3D features obtained from compact auxiliary frames, allow us to synthesize high fidelity temporary reconstructed (TR) frames even under challenging motion. 
Second, we integrate audio signals directly into the reconstruction pipeline, ensuring tight lip synchronization. By aligning the TR frames with the corresponding audio stream, we not only achieve better perceptual realism but also eliminate the common pitfalls of disjoint audio and video encoding processes.

Extensive experiments demonstrate that our audio-visual driven video codec reduces bitrate by 32\% over HEVC, 22\% over VVC, and 8.5\% over the latest learning-based approache.
Moreover, our codec achieves the highest overall scores across two lip-sync evaluation metrics.
Our findings broaden the scope of research in audio-visual compression and pave the way toward more immersive and interactive communication systems.

In summary, our key contributions are threefold:
\begin{itemize}[noitemsep, topsep=3pt]
    \item \textbf{3D Feature Extraction for Robust Facial Modeling.} We propose a novel framework that captures significant head movement and complex facial orientations by extracting compact 3D features, thereby overcoming the limitations of traditional 2D keypoint-based methods.
    \item \textbf{Audio-Driven Alignment for Tight Lip Synchronization.} 
    Our codec integrates audio signals into the reconstruction pipeline, ensuring precise lip-sync and delivering a more natural viewing experience, even under low-bitrates. 
    \item \textbf{End-to-End Audio-visual Codec.} We jointly present the first approach to optimize video and audio streams within a learning-based codec, achieving superior compression efficiency and delivering more natural videos compared to state-of-the-art codecs.
\end{itemize}

\section{Method}
We divide a video sequence into multiple groups, Groups of Pictures (GOPs), which consist of $N$ frames. Within one GOP, we define the first frame as the keyframe, while the subsequent frames are target frames. 

As shown in Fig.~\ref{fig:architecture}, at the encoder, we compress the keyframe using the VVC intra-setting to maintain high-quality. 
For target frames, we prioritize data compactness by downsampling and compressing using VVC low-delay setting.
These frames serve as auxiliary frames for obtaining 3D features relative to the target frames.

At the decoder, we adopt the two-stage reconstruction processes. 
We note that the high-quality keyframe is directly used as an output and leveraged as a reference for reconstructing target frames.
In the first stage, it reconstructs temporarily target frames. 
Appearance features and motion features are extracted from the keyframe and auxiliary frames, respectively. These features are then utilized to reproduce dynamic information, including head movements and facial expressions. This stage enables the representation of head movements that cannot be fully captured using audio alone.
In the second stage, the target frames are further reconstructed using the audio and the TR frames, ensuring synchronization between the audio and the mouth movements. This stage enables the representation of lip-sync accurate mouth movements, which are difficult to achieve using only visual features optimized for compression efficiency.

\subsection{Stage I: Temporary reconstruction of the target frames}
\noindent
\textbf{Facial Super-Resolution (SR).}
Facial SR adjusts the size of auxiliary frames to align them with the keyframe, utilizing the facial super-resolution model GFPGAN~\cite{wang2021gfpgan}.
It enables not only super-resolution but also efficient facial feature enhancement.

\vspace{\baselineskip} 

\noindent
\textbf{Facial Animator.}
Facial Animator temporarily reconstructs the target frames by utilizing 3D features. 
It consists of an Appearance Feature Extractor, a Motion Feature Extractor, a Warping Module, and a Generator, inspired by LivePortrait~\cite{guo2024liveportrait}. 
We leverage the pretrained weights trained on large-scale data in ~\cite{guo2024liveportrait}.

First, we extract a 3D appearance feature \textit{$f_{key}$} from the keyframe using the Appearance Feature Extractor. We also estimate the 3D implicit keypoints $x_{key}$ and $x_{trg,i}$ of the keyframe and target frame as motion representations, respectively:

\begin{equation}
    \left\{
    \begin{aligned}
        x_{key} &= \textit{S}_{key} \cdot (x_{c,key} R_{key} + \delta_{key}) + T_{key}, \\
        x_{trg,i} &= S_{key} \cdot \frac{S_{trg,i}}{S_{trg,0}} 
        \cdot \bigl( x_{c,key} \bigl(R_{trg,i} R^{-1}_{trg,0} R_{key}\bigr) \bigr) \\
        & \quad + (\delta_{key} + \delta_{trg,i} - \delta_{trg,0}) \\
        & \quad + (T_{key} + T_{trg,i} - T_{trg,0}) + \Delta_{mouth,i}.
    \end{aligned}
    \right\}
    \tag{1}
\end{equation}

\begin{table*}[t]
    \caption{Comparison of the coding performance with SOTAs}
    \centering
    \begin{adjustbox}{width=0.9\textwidth}
        \begin{tabular}{c c c c c c c c}
          \hline
          \ Metrics & HEVC ~\cite{sullivan2012overview} & VVC ~\cite{bross2021overview} & DAC ~\cite{konuko2021ultra} & HDAC ~\cite{konuko2022hybrid} & CFTE ~\cite{chen2022beyond} & RDAC ~\cite{konuko2023predictive} & BiLFAC ~\cite{10992457} \\
          \hline
          LPIPS & -86.70\% & -79.49\% & -58.17\% & -23.14\% & -63.86\% & -80.14\% & 6.786\%\\
          DISTS & -99.61\% & -99.92\% & -55.30\% & -40.54\% & -99.24\% & -81.20\% & -8.455\% \\
          FID   & -32.31\% & -22.44\% & -44.84\% & -15.11\% & -57.94\% & -84.37\% & 11.81\%\\
          \hline
        \end{tabular}
    \end{adjustbox}    
    \label{tb:BD-rate}
\end{table*}


\begin{figure*}[t]
    \centering
    \includegraphics[width=\textwidth]{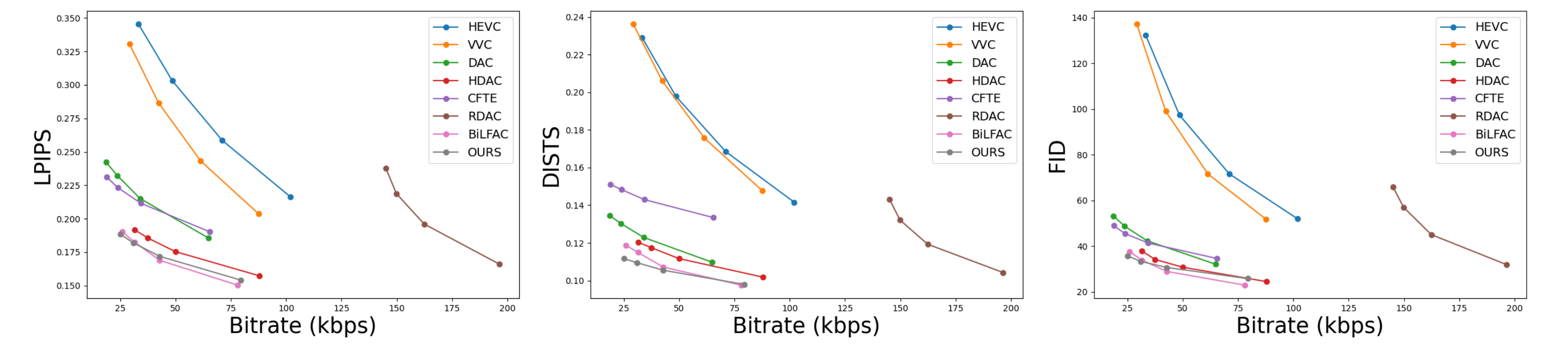}
    \caption{RD performance in terms of LPIPS (left), DISTS (center), and FID (right).}
    \label{fig:rd_comparison}
\end{figure*}

where ${S}$ represents the scale factor, $x_{c,key} \in \mathbb{R}^{K\times3}$ is canonical keypoints of keyframe, $R \in \mathbb{R}^{3\times3}$ is head pose, $\delta \in \mathbb{R}^{K\times3}$ is expression deformation, and $T \in \mathbb{R}^3$ is translation.
These motion features are extracted from keyframe and auxiliary frames by the Motion Extractor.
Additionally, to unify the input conditions for the subsequent processing of Stage I\hspace{-1.2pt}I, we adopt the mouth retargeting module ~\cite{guo2024liveportrait}. We calculate the deformation offset $\Delta_{mouth}$ by the mouth retargeting module and add to $x_{trg}$.
Subsequently, the Facial Animator outputs TR frames.
\vspace{-0.5em} 
\begin{equation}
TR frame_i = G(W(f_{key};x_{key},x_{trg,i})) \tag{2}
\end{equation}

where \textit{W} represents the Warping Module and \textit{G} represents the Generator.

\subsection{Stage I\hspace{-1.2pt}I: Audio and mouth synchronization}
As shown in Fig.~\ref{fig:architecture}, we incorporate the Lip Syncer to reconstruct the final target frames with synchronized audio and lip movements. It consists of a VAE (Variational Autoencoder) ~\cite{kingma2013auto} encoder, an Audio Encoder, an UNet, and a VAE Decoder, inspired by MuseTalk ~\cite{zhang2024musetalk}.

First, we use a VAE Encoder to extract the visual feature $v_{ref}$ from the ${TR frame_i}$. 
In addition, we obtain the masked visual feature $v_{mask}$ from the same ${TR frame_i}$ by masking the lower half of the frame. 
For the audio signal, we resample it to 16,000 Hz and convert it into an 80-dimensional log-mel spectrogram. 
We adopted the Whisper ~\cite{radford2023robust} encoder as the Audio Encoder to extract audio features ${a^{t \times d}}$, where $t$ represents the time and $d$ denotes the dimensionality of the audio features.
For each ${TR frame_i}$, the corresponding audio features are obtained by concatenating the features in the range ${2i -4 \leq t \leq 2i+5}$. 
The visual features $v_{ref}$ and $v_{mask}$ are concatenated along the channel dimension and fed into the UNet ~\cite{ronneberger2015u}. 
Additionally, we feed the audio features into the cross-attention layer, which fuses the visual and audio features into a single representation.
This fused feature is passed to the VAE Decoder, reconstructing the final target frame. 
By leveraging the original ${TR frame_i}$ as a reference, our approach effectively suppresses unexpected deformations of the jawline while ensuring synchronized and natural mouth movements.

\subsection{Loss function and training}
For the Lip Syncer, we leverage the pretrained weights from \\ MuseTalk~\cite{zhang2024musetalk} as a foundation and further refine them through fine-tuning:
\vspace{-4pt}
\[
\mathcal{L}_{\text{rec}} = \frac{1}{N} \sum_{i=1}^N \| I_r^i - I_{\text{gt}}^i \|_1 \hspace{4pt}(3), \hspace{4pt}
\mathcal{L}_{\text{p}} = \frac{1}{N} \sum_{i=1}^N \| \mathcal{V}(I_r^i) - \mathcal{V}(I_{\text{gt}}^i) \|_2 \hspace{4pt} (4)
\]

\begin{equation}
\mathcal{L}_{\text{sync}} = \frac{1}{N} \sum_{i}^{N} -\log \big[P_{\text{sync}}(\text{SyncNet}(A_{mel}^i, I_r^i))\big] \tag{5}
\end{equation}

\noindent
The reconstruction loss \(\mathcal{L}_{\text{rec}}\) and the perceptual loss \(\mathcal{L}_{\text{p}}\) are defined in Equations (3) and (4), respectively.
\(I_r\) represents the reconstructed image, and \(I_{gt}\) represents the ground truth image.
\(\mathcal{V}\) denotes the feature extractor of VGG19 ~\cite{DBLP:journals/corr/SimonyanZ14a}. 
As shown in Equation (5), we introduce a lip-sync loss to synchronize audio and mouth movements by calculating the cosine similarity using pairs of audio and image frames as input to SyncNet ~\cite{chung2017out}.

\vspace{-4pt}

\begin{equation}
\mathcal{L} = \mathcal{L}_{\text{rec}} + \lambda \mathcal{L}_{\text{p}} + \mu \mathcal{L}_{\text{sync}} \tag{6}
\end{equation}

\noindent
Finally, as shown in Equation (6), the final loss is defined as a weighted sum, where we set \(\lambda = 0.01\) and \(\mu = 0.03\) in our experiments.

\begin{figure*}[t]
    \centering
    \includegraphics[width=\textwidth]{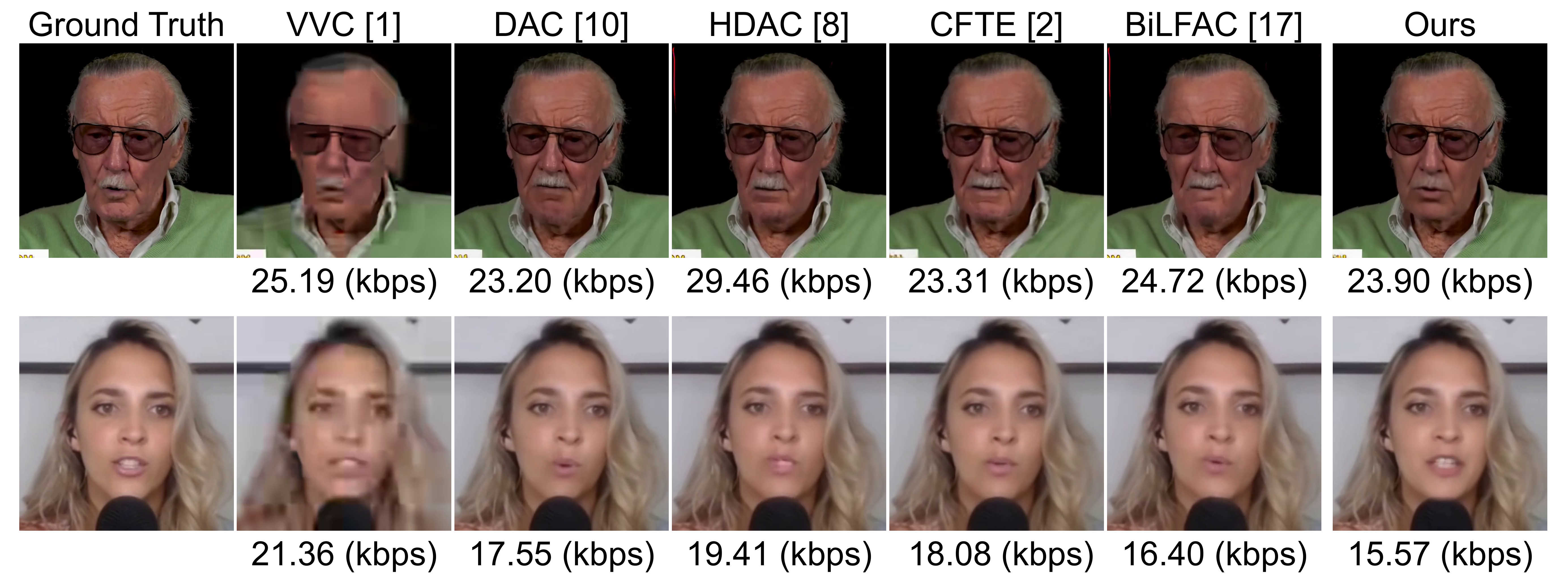}
    \caption{Visual comparison of coding results at comparable bitrates. Compared to existing codecs, our codec achieves better performance by minimizing artifacts and distortions around the face while accurately representing mouth shapes.}
    \label{fig:visual_comparison}
\end{figure*}

\section{Results}

\subsection{Evaluation Protocol}
We use the high-quality video dataset CelebV-HQ~\cite{zhu2022celebv}. All videos were standardized to a frame rate of 25 fps and a frame size of 512×512. We randomly selected 65 videos and tested 180 frames from each video. 
We note that changing the GOP size affects the bitrate and the reconstruction quality. Therefore, in this study, we set the GOP size to 15, 30, 45, and 60 for the test. The quantization parameter (QP) for keyframes was also set to 30. We applied the same QP setting to keyframes in existing learning-based codecs. Furthermore, following previous studies ~\cite{konuko2023predictive, 10992457}, we assigned higher QP values to HEVC and VVC with low-delay configurations to achieve a comparable bitrate.

To evaluate visual quality, we adopt LPIPS~\cite{zhang2018unreasonable}, DISTS~\cite{ding2020image}, and FID~\cite{heusel2017gans}. These visual metrics indicate that a lower score corresponds to higher quality. For lip-sync evaluation, we used LSE-C ~\cite{prajwal2020lip} calculated by SyncNet ~\cite{chung2017out} and AVS\_u ~\cite{yaman2024audio} calculated by AV-HuBERT ~\cite{shi2022learning}. Additionally, these lip-sync metrics indicate that a higher score corresponds to better synchronization with the audio.

\begin{table}
    \caption{Comparison of Lip-Sync with SOTAs at the same bitrate. While \colorbox{red!20}{red} indicates the best score, \colorbox{blue!20}{blue} shows the second best.}
    \centering
    \renewcommand{\arraystretch}{1.0}
    \begin{adjustbox}{width=0.8\linewidth}
        \begin{tabular}{c c c}
          \hline
          Method & LSE\_C $\uparrow$ & AVS\_u $\uparrow$ \\
          \hline
          DAC  ~\cite{konuko2021ultra} & 1.47 & 0.572 \\
          HDAC ~\cite{konuko2022hybrid} & \cellcolor{blue!20}3.51 & 0.559 \\  
          CFTE ~\cite{chen2022beyond} & 1.27 & \cellcolor{red!20}0.608  \\  
          BiLFAC ~\cite{10992457} & 2.27 & 0.576  \\
          \hline
          Ours & \cellcolor{red!20}3.88 & \cellcolor{blue!20}0.581  \\  
          \hline
        \end{tabular}
    \end{adjustbox}    
    \label{tb:lip-sync}
    \vspace{-0.7em}
\end{table}

\subsection{Quantitative Results}
Table~\ref{tb:BD-rate} shows the bitrate savings compared to state-of-the-art codecs (SOTAs). Our codec achieves a 33\% bitrate reduction in FID compared to HEVC and a 22\% reduction compared to VVC.
Furthermore, it reduces the bitrate in most cases when compared to existing learning-based video codecs~\cite{konuko2021ultra, konuko2022hybrid, chen2022beyond, konuko2023predictive, 10992457}.

Fig.~\ref{fig:rd_comparison} provides the rate-distortion performance of the proposed codec compared to existing video codecs.
Our codec achieves higher performance at low-bitrates compared to HEVC and VVC.
In addition, it outperforms almost all metrics compared to existing learning-based video codecs.

Table~\ref{tb:lip-sync} presents the comparison of Lip Synchronization with SOTAs. For a fair comparison, RDAC ~\cite{konuko2023predictive} is not included in Table~\ref{tb:lip-sync} as it failed to achieve an equivalent bitrate. Furthermore, at lower bitrates, HEVC and VVC exhibited block noise, preventing a fair evaluation of lip synchronization; therefore, they were also excluded from Table~\ref{tb:lip-sync}.
The proposed method achieves the highest score in LSE\_C, outperforming all other methods.
It also achieves the second-best score in AVS\_u. These results demonstrate its overall superior lip synchronization performance.

\begin{figure}
    \centering
    \includegraphics[width=\linewidth]{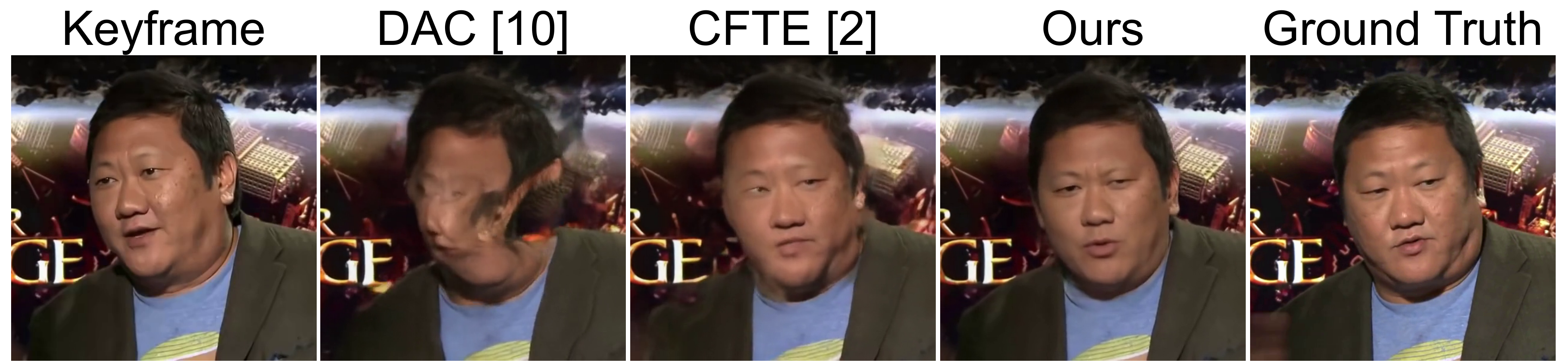}
    \caption{Comparison of target frame reconstruction under large head motion scenarios. Our 3D-based codec effectively represents head movements, while existing 2D-based codecs fail to do so, resulting in facial distortions.}
    \label{fig:frame_comparison}
    \vspace{-1.3em}
\end{figure}

\subsection{Qualitative Results}
As shown in Fig.~\ref{fig:visual_comparison}, our codec achieves the highest subjective quality compared to others at a similar bitrate.
Existing codecs introduce distortions around the mouth and fail to allow the mouth to open fully.
In contrast, our codec reproduces mouth shapes closest to the ground truth among all codecs. It indicates that leveraging audio for reconstruction ensures accurate synchronization between the audio and mouth movements.

As shown in Fig.~\ref{fig:frame_comparison}, when reconstructing a frame where a person is facing forward from a keyframe where the person is facing sideways, existing codecs using 2D features produce distortions around the face. In contrast, our codec, which utilizes 3D features, accurately captures perspective and depth information, enabling frame reconstruction without any distortions around the face.

\section{Conclusion}
In this paper, we propose an audio-visual driven video codec that addresses the dual challenges of achieving precise lip synchronization and handling large head movements in bandwidth-limited scenarios. 
Our 3D feature-based framework integrates audio features, enabling robust facial motion modeling and natural lip movements synchronized with audio.
Experiments on the CelebV-HQ dataset show that our method reduces bitrate by up to 33\% compared to HEVC and 22\% relative to VVC, outperforming state-of-the-art learning-based codecs in both visual quality and lip-sync accuracy. 
Qualitative evaluations demonstrate that, even under challenging pose variations, our method accurately captures facial movements and represents precise lip movements compared to existing codecs. 

\begin{acks}
This work was supported by JSPS KAKENHI under Grant No. JP25K15165.
\end{acks}


\begin{thebibliography}{22}


\ifx \showCODEN    \undefined \def \showCODEN     #1{\unskip}     \fi
\ifx \showISBNx    \undefined \def \showISBNx     #1{\unskip}     \fi
\ifx \showISBNxiii \undefined \def \showISBNxiii  #1{\unskip}     \fi
\ifx \showISSN     \undefined \def \showISSN      #1{\unskip}     \fi
\ifx \showLCCN     \undefined \def \showLCCN      #1{\unskip}     \fi
\ifx \shownote     \undefined \def \shownote      #1{#1}          \fi
\ifx \showarticletitle \undefined \def \showarticletitle #1{#1}   \fi
\ifx \showURL      \undefined \def \showURL       {\relax}        \fi
\providecommand\bibfield[2]{#2}
\providecommand\bibinfo[2]{#2}
\providecommand\natexlab[1]{#1}
\providecommand\showeprint[2][]{arXiv:#2}

\bibitem[Bross et~al\mbox{.}(2021)]%
        {bross2021overview}
\bibfield{author}{\bibinfo{person}{Benjamin Bross}, \bibinfo{person}{Ye-Kui Wang}, \bibinfo{person}{Yan Ye}, \bibinfo{person}{Shan Liu}, \bibinfo{person}{Jianle Chen}, \bibinfo{person}{Gary~J Sullivan}, {and} \bibinfo{person}{Jens-Rainer Ohm}.} \bibinfo{year}{2021}\natexlab{}.
\newblock \showarticletitle{Overview of the versatile video coding (VVC) standard and its applications}.
\newblock \bibinfo{journal}{\emph{IEEE Transactions on Circuits and Systems for Video Technology}} \bibinfo{volume}{31}, \bibinfo{number}{10} (\bibinfo{year}{2021}), \bibinfo{pages}{3736--3764}.
\newblock


\bibitem[Chen et~al\mbox{.}(2022)]%
        {chen2022beyond}
\bibfield{author}{\bibinfo{person}{Bolin Chen}, \bibinfo{person}{Zhao Wang}, \bibinfo{person}{Binzhe Li}, \bibinfo{person}{Rongqun Lin}, \bibinfo{person}{Shiqi Wang}, {and} \bibinfo{person}{Yan Ye}.} \bibinfo{year}{2022}\natexlab{}.
\newblock \showarticletitle{Beyond keypoint coding: Temporal evolution inference with compact feature representation for talking face video compression}. In \bibinfo{booktitle}{\emph{2022 Data Compression Conference (DCC)}}. IEEE, \bibinfo{pages}{13--22}.
\newblock


\bibitem[Chung and Zisserman(2017)]%
        {chung2017out}
\bibfield{author}{\bibinfo{person}{Joon~Son Chung} {and} \bibinfo{person}{Andrew Zisserman}.} \bibinfo{year}{2017}\natexlab{}.
\newblock \showarticletitle{Out of time: automated lip sync in the wild}. In \bibinfo{booktitle}{\emph{Computer Vision--ACCV 2016 Workshops: ACCV 2016 International Workshops, Taipei, Taiwan, November 20-24, 2016, Revised Selected Papers, Part II 13}}. Springer, \bibinfo{pages}{251--263}.
\newblock


\bibitem[Ding et~al\mbox{.}(2020)]%
        {ding2020image}
\bibfield{author}{\bibinfo{person}{Keyan Ding}, \bibinfo{person}{Kede Ma}, \bibinfo{person}{Shiqi Wang}, {and} \bibinfo{person}{Eero~P Simoncelli}.} \bibinfo{year}{2020}\natexlab{}.
\newblock \showarticletitle{Image quality assessment: Unifying structure and texture similarity}.
\newblock \bibinfo{journal}{\emph{IEEE transactions on pattern analysis and machine intelligence}} \bibinfo{volume}{44}, \bibinfo{number}{5} (\bibinfo{year}{2020}), \bibinfo{pages}{2567--2581}.
\newblock


\bibitem[Guo et~al\mbox{.}(2024)]%
        {guo2024liveportrait}
\bibfield{author}{\bibinfo{person}{Jianzhu Guo}, \bibinfo{person}{Dingyun Zhang}, \bibinfo{person}{Xiaoqiang Liu}, \bibinfo{person}{Zhizhou Zhong}, \bibinfo{person}{Yuan Zhang}, \bibinfo{person}{Pengfei Wan}, {and} \bibinfo{person}{Di Zhang}.} \bibinfo{year}{2024}\natexlab{}.
\newblock \showarticletitle{LivePortrait: Efficient Portrait Animation with Stitching and Retargeting Control}.
\newblock \bibinfo{journal}{\emph{arXiv preprint arXiv:2407.03168}} (\bibinfo{year}{2024}).
\newblock


\bibitem[Heusel et~al\mbox{.}(2017)]%
        {heusel2017gans}
\bibfield{author}{\bibinfo{person}{Martin Heusel}, \bibinfo{person}{Hubert Ramsauer}, \bibinfo{person}{Thomas Unterthiner}, \bibinfo{person}{Bernhard Nessler}, {and} \bibinfo{person}{Sepp Hochreiter}.} \bibinfo{year}{2017}\natexlab{}.
\newblock \showarticletitle{Gans trained by a two time-scale update rule converge to a local nash equilibrium}.
\newblock \bibinfo{journal}{\emph{Advances in neural information processing systems}}  \bibinfo{volume}{30} (\bibinfo{year}{2017}).
\newblock


\bibitem[Kingma(2013)]%
        {kingma2013auto}
\bibfield{author}{\bibinfo{person}{Diederik~P Kingma}.} \bibinfo{year}{2013}\natexlab{}.
\newblock \showarticletitle{Auto-encoding variational bayes}.
\newblock \bibinfo{journal}{\emph{arXiv preprint arXiv:1312.6114}} (\bibinfo{year}{2013}).
\newblock


\bibitem[Konuko et~al\mbox{.}(2022)]%
        {konuko2022hybrid}
\bibfield{author}{\bibinfo{person}{Goluck Konuko}, \bibinfo{person}{St{\'e}phane Lathuili{\`e}re}, {and} \bibinfo{person}{Giuseppe Valenzise}.} \bibinfo{year}{2022}\natexlab{}.
\newblock \showarticletitle{A hybrid deep animation codec for low-bitrate video conferencing}. In \bibinfo{booktitle}{\emph{2022 IEEE International Conference on Image Processing (ICIP)}}. IEEE, \bibinfo{pages}{1--5}.
\newblock


\bibitem[Konuko et~al\mbox{.}(2023)]%
        {konuko2023predictive}
\bibfield{author}{\bibinfo{person}{Goluck Konuko}, \bibinfo{person}{St{\'e}phane Lathuili{\`e}re}, {and} \bibinfo{person}{Giuseppe Valenzise}.} \bibinfo{year}{2023}\natexlab{}.
\newblock \showarticletitle{Predictive coding for animation-based video compression}. In \bibinfo{booktitle}{\emph{2023 IEEE International Conference on Image Processing (ICIP)}}. IEEE, \bibinfo{pages}{2810--2814}.
\newblock


\bibitem[Konuko et~al\mbox{.}(2021)]%
        {konuko2021ultra}
\bibfield{author}{\bibinfo{person}{Goluck Konuko}, \bibinfo{person}{Giuseppe Valenzise}, {and} \bibinfo{person}{St{\'e}phane Lathuili{\`e}re}.} \bibinfo{year}{2021}\natexlab{}.
\newblock \showarticletitle{Ultra-low bitrate video conferencing using deep image animation}. In \bibinfo{booktitle}{\emph{ICASSP 2021-2021 IEEE International Conference on Acoustics, Speech and Signal Processing (ICASSP)}}. IEEE, \bibinfo{pages}{4210--4214}.
\newblock


\bibitem[Prajwal et~al\mbox{.}(2020)]%
        {prajwal2020lip}
\bibfield{author}{\bibinfo{person}{KR Prajwal}, \bibinfo{person}{Rudrabha Mukhopadhyay}, \bibinfo{person}{Vinay~P Namboodiri}, {and} \bibinfo{person}{CV Jawahar}.} \bibinfo{year}{2020}\natexlab{}.
\newblock \showarticletitle{A lip sync expert is all you need for speech to lip generation in the wild}. In \bibinfo{booktitle}{\emph{Proceedings of the 28th ACM international conference on multimedia}}. \bibinfo{pages}{484--492}.
\newblock


\bibitem[Radford et~al\mbox{.}(2023)]%
        {radford2023robust}
\bibfield{author}{\bibinfo{person}{Alec Radford}, \bibinfo{person}{Jong~Wook Kim}, \bibinfo{person}{Tao Xu}, \bibinfo{person}{Greg Brockman}, \bibinfo{person}{Christine McLeavey}, {and} \bibinfo{person}{Ilya Sutskever}.} \bibinfo{year}{2023}\natexlab{}.
\newblock \showarticletitle{Robust speech recognition via large-scale weak supervision}. In \bibinfo{booktitle}{\emph{International conference on machine learning}}. PMLR, \bibinfo{pages}{28492--28518}.
\newblock


\bibitem[Ronneberger et~al\mbox{.}(2015)]%
        {ronneberger2015u}
\bibfield{author}{\bibinfo{person}{Olaf Ronneberger}, \bibinfo{person}{Philipp Fischer}, {and} \bibinfo{person}{Thomas Brox}.} \bibinfo{year}{2015}\natexlab{}.
\newblock \showarticletitle{U-net: Convolutional networks for biomedical image segmentation}. In \bibinfo{booktitle}{\emph{Medical image computing and computer-assisted intervention--MICCAI 2015: 18th international conference, Munich, Germany, October 5-9, 2015, proceedings, part III 18}}. Springer, \bibinfo{pages}{234--241}.
\newblock


\bibitem[Shi et~al\mbox{.}(2022)]%
        {shi2022learning}
\bibfield{author}{\bibinfo{person}{Bowen Shi}, \bibinfo{person}{Wei-Ning Hsu}, \bibinfo{person}{Kushal Lakhotia}, {and} \bibinfo{person}{Abdelrahman Mohamed}.} \bibinfo{year}{2022}\natexlab{}.
\newblock \showarticletitle{Learning audio-visual speech representation by masked multimodal cluster prediction}.
\newblock \bibinfo{journal}{\emph{arXiv preprint arXiv:2201.02184}} (\bibinfo{year}{2022}).
\newblock


\bibitem[Simonyan and Zisserman(2015)]%
        {DBLP:journals/corr/SimonyanZ14a}
\bibfield{author}{\bibinfo{person}{Karen Simonyan} {and} \bibinfo{person}{Andrew Zisserman}.} \bibinfo{year}{2015}\natexlab{}.
\newblock \showarticletitle{Very Deep Convolutional Networks for Large-Scale Image Recognition}. In \bibinfo{booktitle}{\emph{3rd International Conference on Learning Representations, {ICLR} 2015, San Diego, CA, USA, May 7-9, 2015, Conference Track Proceedings}}, \bibfield{editor}{\bibinfo{person}{Yoshua Bengio} {and} \bibinfo{person}{Yann LeCun}} (Eds.).
\newblock
\urldef\tempurl%
\url{http://arxiv.org/abs/1409.1556}
\showURL{%
\tempurl}


\bibitem[Sullivan et~al\mbox{.}(2012)]%
        {sullivan2012overview}
\bibfield{author}{\bibinfo{person}{Gary~J Sullivan}, \bibinfo{person}{Jens-Rainer Ohm}, \bibinfo{person}{Woo-Jin Han}, {and} \bibinfo{person}{Thomas Wiegand}.} \bibinfo{year}{2012}\natexlab{}.
\newblock \showarticletitle{Overview of the high efficiency video coding (HEVC) standard}.
\newblock \bibinfo{journal}{\emph{IEEE Transactions on circuits and systems for video technology}} \bibinfo{volume}{22}, \bibinfo{number}{12} (\bibinfo{year}{2012}), \bibinfo{pages}{1649--1668}.
\newblock


\bibitem[Takahashi et~al\mbox{.}(2025)]%
        {10992457}
\bibfield{author}{\bibinfo{person}{Riku Takahashi}, \bibinfo{person}{Ryugo Morita}, \bibinfo{person}{Fuma Kimishima}, \bibinfo{person}{Kosuke Iwama}, {and} \bibinfo{person}{Jinjia Zhou}.} \bibinfo{year}{2025}\natexlab{}.
\newblock \showarticletitle{Bidirectional Learned Facial Animation Codec for Low Bitrate Talking Head Videos}. In \bibinfo{booktitle}{\emph{2025 Data Compression Conference (DCC)}}. \bibinfo{pages}{401--401}.
\newblock
\href{https://doi.org/10.1109/DCC62719.2025.00088}{doi:\nolinkurl{10.1109/DCC62719.2025.00088}}


\bibitem[Wang et~al\mbox{.}(2021)]%
        {wang2021gfpgan}
\bibfield{author}{\bibinfo{person}{Xintao Wang}, \bibinfo{person}{Yu Li}, \bibinfo{person}{Honglun Zhang}, {and} \bibinfo{person}{Ying Shan}.} \bibinfo{year}{2021}\natexlab{}.
\newblock \showarticletitle{Towards Real-World Blind Face Restoration with Generative Facial Prior}. In \bibinfo{booktitle}{\emph{The IEEE Conference on Computer Vision and Pattern Recognition (CVPR)}}.
\newblock


\bibitem[Yaman et~al\mbox{.}(2024)]%
        {yaman2024audio}
\bibfield{author}{\bibinfo{person}{Dogucan Yaman}, \bibinfo{person}{Fevziye~Irem Eyiokur}, \bibinfo{person}{Leonard B{\"a}rmann}, \bibinfo{person}{Seymanur Akti}, \bibinfo{person}{Haz{\i}m~Kemal Ekenel}, {and} \bibinfo{person}{Alexander Waibel}.} \bibinfo{year}{2024}\natexlab{}.
\newblock \showarticletitle{Audio-Visual Speech Representation Expert for Enhanced Talking Face Video Generation and Evaluation}. In \bibinfo{booktitle}{\emph{Proceedings of the IEEE/CVF Conference on Computer Vision and Pattern Recognition}}. \bibinfo{pages}{6003--6013}.
\newblock


\bibitem[Zhang et~al\mbox{.}(2018)]%
        {zhang2018unreasonable}
\bibfield{author}{\bibinfo{person}{Richard Zhang}, \bibinfo{person}{Phillip Isola}, \bibinfo{person}{Alexei~A Efros}, \bibinfo{person}{Eli Shechtman}, {and} \bibinfo{person}{Oliver Wang}.} \bibinfo{year}{2018}\natexlab{}.
\newblock \showarticletitle{The unreasonable effectiveness of deep features as a perceptual metric}. In \bibinfo{booktitle}{\emph{Proceedings of the IEEE conference on computer vision and pattern recognition}}. \bibinfo{pages}{586--595}.
\newblock


\bibitem[Zhang et~al\mbox{.}(2024)]%
        {zhang2024musetalk}
\bibfield{author}{\bibinfo{person}{Yue Zhang}, \bibinfo{person}{Minhao Liu}, \bibinfo{person}{Zhaokang Chen}, \bibinfo{person}{Bin Wu}, \bibinfo{person}{Yubin Zeng}, \bibinfo{person}{Chao Zhan}, \bibinfo{person}{Yingjie He}, \bibinfo{person}{Junxin Huang}, {and} \bibinfo{person}{Wenjiang Zhou}.} \bibinfo{year}{2024}\natexlab{}.
\newblock \showarticletitle{Musetalk: Real-time high quality lip synchronization with latent space inpainting}.
\newblock \bibinfo{journal}{\emph{arXiv preprint arXiv:2410.10122}} (\bibinfo{year}{2024}).
\newblock


\bibitem[Zhu et~al\mbox{.}(2022)]%
        {zhu2022celebv}
\bibfield{author}{\bibinfo{person}{Hao Zhu}, \bibinfo{person}{Wayne Wu}, \bibinfo{person}{Wentao Zhu}, \bibinfo{person}{Liming Jiang}, \bibinfo{person}{Siwei Tang}, \bibinfo{person}{Li Zhang}, \bibinfo{person}{Ziwei Liu}, {and} \bibinfo{person}{Chen~Change Loy}.} \bibinfo{year}{2022}\natexlab{}.
\newblock \showarticletitle{CelebV-HQ: A large-scale video facial attributes dataset}. In \bibinfo{booktitle}{\emph{European conference on computer vision}}. Springer, \bibinfo{pages}{650--667}.
\newblock


\end{thebibliography}

\end{document}